\documentclass[11pt,a4paper]{article}
\pdfoutput=1 % if your are submitting a pdflatex (i.e. if you have
\usepackage{jcappub} % for details on the use of the package, please
\usepackage[T1]{fontenc} % if needed

\title{\boldmath Asymmetric dark matter\\
in braneworld cosmology}

\author{Michael T. Meehan}
\author{and Ian B. Whittingham}
\affiliation{James Cook University,\\ 1 James Cook Dr.\\ QLD, 4811, Australia}

\emailAdd{Michael.Meehan@my.jcu.edu.au}
\emailAdd{Ian.Whittingham@jcu.edu.au}

\abstract{We investigate the effect of a braneworld expansion era on the relic density
of asymmetric dark matter. We find that the enhanced expansion rate in the early universe 
predicted by the Randall-Sundrum II (RSII) model leads to earlier particle freeze-out 
and an enhanced relic density. This effect has been observed previously by 
Okada and Seto (2004) for symmetric dark matter models and here we extend their 
results to the case of asymmetric dark matter. We also discuss the enhanced asymmetric 
annihilation rate in the braneworld scenario and its implications for indirect 
detection experiments.}

\begin{document}
\maketitle
\flushbottom

\section{Introduction}
\label{sec:intro}

Despite the overwhelming astrophysical and cosmological evidence for the existence of 
Dark Matter (DM)~\cite{Hooper1}, very little is known about its particle nature. 
Many particle candidates have been proposed which are capable of explaining the 
observational data yet none have been conclusively verified. 
The data favour cold (non-relativistic) DM for which the most popular theoretical 
candidates are WIMPs (Weakly Interacting Massive Particles) with mass 
$m_{\chi} \sim \mathcal{O}(10 - 1000)$ GeV. Supersymmetric extensions of the 
Standard Model (SM) in which $R$-parity is conserved provide a viable DM candidate, 
the neutralino, which is the lightest supersymmetric particle formed from 
higgsinos and weak gauginos and is stable against decay into SM particles.

A popular framework for the origin of dark matter is provided by the thermal relic 
scenario; at early times, when the temperature of the universe is high, 
frequent interactions keep the dark matter (anti) particles, $(\bar{\chi})\chi$, 
in equilibrium with the background cosmic bath. As the universe expands and cools 
the dark matter interaction rate drops below the expansion rate and 
the particles fall out of equilibrium. Eventually both creation and annihilation 
processes cease, and the number density redshifts with the expansion. 
This is known as particle freeze-out, and the remaining 'relic' particles constitute 
the dark matter density we observe today. 

A combined analysis of the \textit{Planck} satellite + WP + highL + BAO measurements 
gives the present dark matter density as ($68\%$ C.L.)~\cite{Lahav}
\begin{equation}
\Omega_{DM}h^2 = 0.1187\pm 0.0017,\label{eq:dm_abun}
\end{equation}
where $h = 0.678\pm 0.008$ and is defined by the value of the Hubble constant, 
$H_0 = 100\,h$ km/s/Mpc. 

Due to the Boltzmann suppression factor in the equilibrium number density, the 
longer the DM (anti) particles remain in equilibrium the lower their number 
densities are at freeze-out. Thus species with larger interaction cross sections 
which maintain thermal contact longer, freeze out with diminished abundances. 
Thermal relic WIMPS are excellent DM candidates as their weak scale cross section 
$\sigma \sim G_{\mathrm{F}}^{2}m_{\chi}^{2}$ gives the correct order of magnitude for
$\Omega_{DM}h^{2}$ for a standard radiation-dominated early universe.
However, if the universe experiences a non-standard expansion law during 
the epoch of dark matter decoupling, freeze-out may be accelerated and the relic 
abundance enhanced~\cite{Catena,Barrow}. 

The physics of the early universe, prior to the era of Big Bang Nucleosynthesis (BBN), is relatively unconstrained by current observational datasets. Dark matter particles, which decouple from the background thermal bath at early times, carry the signature of these earliest moments and therefore provide an excellent observational probe. If the properties of the dark matter particles are ever discovered, either through direct/indirect detection experiments or via particle creation in particle accelerators~\cite{Bauer}, then relic abundance calculations could provide a valuable insight into the conditions of the universe prior to BBN. 

The majority of dark matter models assume symmetric dark matter for which the particles 
are Majorana fermions with $\chi = \bar{\chi}$, i.e. they are self-conjugate. 
Given that most known particles are not Majorana, it is natural to consider asymmetric dark matter models in which the particle $\chi$ and antiparticle $\bar{\chi}$ are distinct, i.e. 
$\chi\neq\bar{\chi}$, and to assume an asymmetry between the number densities of the 
DM particles and antiparticles. 
Indeed, a similar asymmetry exists in the baryonic matter sector between 
the number of observed baryons $n_{b}$ and antibaryons $n_{\bar{b}}$. 
This baryonic asymmetry is
\begin{equation}
\eta_{b} = \frac{n_{B}}{n_{\gamma}} = \frac{n_b - n_{\bar{b}}}{n_\gamma} \approx 6\times 10^{-10},
\end{equation}
where $n_{\gamma}$ is the number density of photons. 
Several models have been proposed~\cite{Kumar} that relate the asymmetries in the baryonic
and dark sectors. These models typically assume~\cite{Graesser} either a primordial 
asymmetry in one sector which is transferred to the other sector, or that both
asymmetries are generated by the same physical process such as the decay of a heavy particle.
Kaplan et al.~\cite{Kaplan} consider a baryonic $B-L$ asymmetry generated by 
baryogenesis at high temperatures that is transferred to the DM sector by interactions
arising from higher dimension operators which
then decouple at a temperature above the DM mass and freeze in the asymmetry.
If the asymmetries in the dark and baryonic matter sectors share a common origin 
then their number densities will be related $n_{DM}\sim n_b$, as will their densities 
$\Omega_{DM}\sim (m_{\chi}/m_b)\Omega_b$~\cite{Kaplan}. This could explain the approximate 
equality of the observed dark and baryonic abundances ($\Omega_{DM}/\Omega_b \sim 5$) 
and suggests a WIMP mass in the range $m_\chi \sim 5 - 15$ GeV. Interestingly, 
this mass range is favored by a number of observational datasets~\cite{DAMA,CoGeNT,Hooper2} 
providing further motivation for asymmetric DM.

Cosmological, astrophysical and collider constraints on light thermal DM ($m_{\chi} \sim
$1 MeV - 10 GeV) have been examined by~\cite{Lin} for both symmetric and asymmetric 
models of DM and~\cite{Kim} have considered flavour constraints on, and collider 
signatures of, asymmetric DM produced by decays of supersymmetric particles in 
the minimal supersymmetric Standard Model.

The relic abundance of asymmetric DM has been studied~\cite{Gelmini,Iminniyaz1,Iminniyaz2} in 
the standard cosmological scenario and for the non-standard quintessence scenario in 
which a non-interacting scalar field is present in  its kination phase. Gelmini 
et al.~\cite{Gelmini} also considered a simple scalar-tensor cosmology 
parameterized as a multiplicatively modified Hubble expansion. 
The enhanced expansion rate predicted by each non-standard scenario led to earlier particle 
freeze-out and an enhanced relic abundance. As a result, the asymmetry between the particles and antiparticles was essentially 'washed out'.

In this study we consider the effect of an early time braneworld expansion era on 
the present density of asymmetric dark matter. Braneworld models are toy models 
arising from string theory where additional spacetime dimensions are incorporated in an attempt to unify the fundamental forces of nature.
In the braneworld scenario, our universe is modeled as a $3(+1)$ dimensional surface (the brane) embedded in a higher dimensional spacetime known as the bulk. The standard model particles are confined to the surface of the brane whilst gravity resides in the higher dimensional bulk. This offers an explanation for the apparent weakness of gravity with respect to the other fundamental forces~\cite{Langlois}. 

The effect of a braneworld expansion era on the relic abundance of symmetric DM has 
been studied by~\cite{Okada1,Nihei1,Nihei2,Dahib, Guo}. They found that the modified expansion rate 
in braneworld models led to earlier particle freeze-out and an enhanced relic abundance 
(provided the five-dimensional Planck mass $M_{5}$ is low enough),
similar features to those of the quintessence and scalar-tensor scenarios. We report here an extension of these studies to the case of asymmetric DM.  

In the next section we introduce the Randall-Sundrum type II braneworld model and its relevant parameters. Then in section~\ref{sec:ADM} we present the Boltzmann equations which describe the time evolution of the asymmetric DM number densities and give both numerical and 
analytical solutions for the braneworld case. We constrain the possible parameter 
combinations using the observed DM density~\eqref{eq:dm_abun} in section~\ref{sec:parbound} 
before discussing the 
asymmetric DM annihilation rate and prospects for indirect detection of asymmetric DM 
in sections~\ref{sec:annrate} and~\ref{sec:Fermi} respectively. 
Finally we summarize our findings in section~\ref{sec:concl}.

\section{Braneworld model}

\label{sec:brane}

In this work we focus on the Randall-Sundrum II (RSII) model~\cite{Randall} in which our universe is realized on a $3( + 1)$-Minkowski brane with positive tension, located at the ultraviolet boundary of the five dimensional anti-de Sitter bulk with cosmological constant $\Lambda_{5} < 0$. 
The expansion rate in this model is given by the modified Friedmann equation,
\begin{equation}
H^2 = \frac{8\pi}{3 M_{\mathrm{Pl}}^2}\rho\left(1 + \frac{\rho}{2\sigma}\right),\label{eq:brane_hub}
\end{equation}
where $\sigma = 48\pi M_5^6/M_{\mathrm{Pl}}^2$ is the brane tension, the four-dimensional 
cosmological constant has been fine-tuned to zero and $M_5$ is the five dimensional 
Planck mass which we will treat as a free parameter. The four dimensional Planck mass 
$M_{\mathrm{Pl}} = 1.22\times 10^{19}$ GeV is related to Newton's constant by 
$G = M_{\mathrm{Pl}}^{-2}$. We have omitted the so-called 'dark radiation' term 
$\propto a^{-4}$ (where $a$ is the cosmic scale factor) because of the severe constraints
on extra light degrees of freedom at the time of Big Bang Nucleosynthesis~\cite{Ichiki}. 

A novel feature of~\eqref{eq:brane_hub} is the presence of the term quadratic in $\rho$. At high energies ($\rho\gg 2\sigma$) this term dominates the expansion and $H\sim \rho$. Comparing this to the standard scenario where $H_{ST}\sim\rho^{1/2}$ we see that the early time expansion rate is enhanced in the RSII braneworld model. As the energy density drops ($\rho \ll 2\sigma$) the quadratic term becomes negligible and the standard expansion law is recovered. 

Since the early universe is radiation dominated, the energy density is 
$\rho \simeq \rho_r = \pi^2 g_{\star}(T)T^4/30$, where $g_\star(T)$ counts the number of 
relativistic degrees of freedom at temperature $T$ \footnote{Since the number of relativistic degrees of freedom $g_{\star\rho}$ only differs from the number of entropic degrees of freedom $g_{\star\,s}$ when the temperature crosses a mass threshold, we take $g_{\star\rho}\simeq g_{\star\,s}= g_\star$ throughout.}, and we can rewrite the modified Friedmann equation as
\begin{equation}
H^2 = H_{ST}^2\left[1 + \left(\frac{x_t}{x}\right)^4\right],\label{eq:hubxt}
\end{equation}
where $H_{ST}^2 = 8\pi\rho/3M_{\mathrm{Pl}}^{2}$ is the expansion rate for the standard 
General Relativity scenario, $x \equiv m_{\chi}/T$ and 
\begin{equation}
x_t^4 = g_\star\,\frac{\pi}{2880}\frac{m_\chi^4}{M_5^6}M_{\mathrm{Pl}}^2 \label{eq:xtdef}.
\end{equation}
The parameter $x_t \equiv m_\chi/T_t$  denotes the transition point from the brane expansion 
era to the standard expansion era. We note that smaller values of $M_5$ give a larger $x_t$ and a greater departure from the standard expansion history.

To preserve the successful predictions of Big Bang Nucleosynthesis we must ensure that the standard expansion rate is restored before $T_{BBN} \simeq 1$ MeV. This translates to a transition point of $x_t \lesssim 10^5\,m_{100}$ where $m_{100}$ is the DM mass in units of 100 GeV. In terms of the parameter $M_5$ we require $M_5 \gtrsim 1.1\times 10^4$ GeV \footnote{More stringent constraints have been placed on $M_5$ from sub-millimeter measurements of the gravitational force and the requirement of
a vanishing cosmological constant, however these constraints are sensitive to the presence of a
bulk scalar field~\cite{Maeda} and will not be considered here.}. 

In order to modify the canonical relic abundance result the DM particles must freeze-out before the standard expansion rate is restored. The standard freeze-out point $x_f\sim 20$ depends only logarithmically on the DM mass~\cite{Kolb} whereas, from (\ref{eq:xtdef}), the transition point, $x_t$, is proportional to the DM mass. The braneworld effects will therefore be more exaggerated for larger masses. Conversely, if the DM mass is small, a larger transition point is required to modify the relic abundance.

\section{Asymmetric dark matter}
\label{sec:ADM}
\subsection{Relic density}
\label{sec:rd}

The relic density of asymmetric DM is obtained by solving the relevant Boltzmann equations for
the particle $\chi$ and antiparticle $\bar{\chi}$. Assuming that the only annihilation processes
are those of $\chi \bar{\chi}$ pairs into Standard Model particles, that is there is no self-annihilation involving $\chi \chi $ or $\bar{\chi} \bar{\chi}$ pairs, the time evolution of 
the number densities $n_{\chi,\bar{\chi}}$ are given by the coupled system of Boltzmann equations
\begin{align}
\frac{dn_\chi}{dt} &=- 3Hn_\chi -\langle\sigma v\rangle\left(n_\chi n_{\bar{\chi}} - n_\chi^{eq}n_{\bar{\chi}}^{eq}\right),\label{eq:nchi} \\
\frac{dn_{\bar{\chi}}}{dt} &= - 3Hn_{\bar{\chi}} -\langle\sigma v\rangle\left(n_\chi n_{\bar{\chi}} - n_\chi^{eq}n_{\bar{\chi}}^{eq}\right),\label{eq:nchibar}
\end{align}
where $H$ is the expansion rate of the universe, $\langle\sigma v\rangle$ is the thermally averaged annihilation cross section multiplied by the relative velocity $v$ of the annihilating
$\chi \bar{\chi}$ pair (loosely termed the "annihilation cross section") and 
$n_{\chi,\bar{\chi}}^{eq}$ are the equilibrium number densities of the particle and antiparticle components. Assuming the DM particles are non-relativistic at
decoupling, the equilibrium densities are 
\begin{align}
n_\chi^{eq} &= g_\chi\left(\frac{m_\chi T}{2\pi}\right)^{3/2}e^{(-m_\chi + \mu_\chi)/T},\\
n_{\bar{\chi}}^{eq} &= g_\chi\left(\frac{m_\chi T}{2\pi}\right)^{3/2}e^{(-m_\chi - \mu_\chi)/T},
\end{align}
where $g_{\chi}$ is the number of internal degrees of freedom of $\chi $ and we have used the fact 
that, in equilibrium, the chemical potentials $\mu_\chi $ and $\mu_{\bar{\chi}}$ satisfy 
$\mu_\chi = -\mu_{\bar{\chi}}$. 

The thermally averaged annihilation cross section is generally parameterized as
\begin{equation}
\langle\sigma v\rangle = a + \frac{bT}{m_\chi},
\end{equation}
where the constant term, $a$, corresponds to $s$-wave scattering and the temperature dependent term, $b$, to $p$-wave scattering. Throughout this work we assume that annihilations are dominated by $s-$wave processes and set $b=0$.

It is convenient to rewrite the Boltzmann equations in terms of the dimensionless variables 
$x = m_\chi/T$ and $Y_{\chi,\bar{\chi}} = n_{\chi,\bar{\chi}}/s$, where $s = 2\pi^2 g_\star(T) T^3/45$ is the entropy density. The system~\eqref{eq:nchibar} tranforms to
\begin{align}
\frac{dY_{\chi}}{dx}&=-A(x)\left(Y_\chi Y_{\bar{\chi}} - Y_{\chi}^{eq}Y_{\bar{\chi}}^{eq}\right),\nonumber\\
\frac{dY_{\bar{\chi}}}{dx}&=-A(x)\left(Y_\chi Y_{\bar{\chi}} - Y_{\chi}^{eq}Y_{\bar{\chi}}^{eq}\right),\label{eq:dYasym1}
\end{align}
where the coefficient $A(x)$ is defined as
\begin{equation}
A(x) \equiv \frac{s\langle\sigma v\rangle}{x H}\,\zeta(x),\label{eq:Adef}
\end{equation}
and $\zeta(x)$ is a temperature dependent variable related to the change in the number of degrees of freedom
\begin{equation}
\zeta(x) = 1 - \frac{1}{3}\frac{d\log{g_{\star}}}{d\log{x}}.
\end{equation}
Assuming that the entropy density is conserved (i.e. $sa^3 =$ const.) the variable 
$Y_{\chi(\bar{\chi})}$ may be interpreted as the comoving number density of the 
particle (antiparticle) component\footnote{Many authors neglect the temperature evolution 
of $g_\star(T)$ and set $\zeta(x) = 1$. 
However, as pointed out by~\cite{Steigman}, the temperature 
dependence of these terms may have an appreciable effect on the final relic density. 
For braneworld models in particular, annihilations can persist for an extended period 
after particle freeze-out before the number density reaches its asymptotic value~\cite{Okada1}. 
During this time the value of $g_\star$ may vary by up to an order of magnitude and 
the resulting relic density can be out by a factor of 2 if the full temperature dependence of this term is not maintained. In what follows we only fix the number of 
relativistic degrees of freedom (and in turn set $\zeta(x) = 1$) in order to derive 
analytic solutions to the Boltzmann equations. However, in our numerical analysis, 
the full temperature dependence is maintained.}. 

Since we have assumed that only interactions of the type $\chi\bar{\chi} \leftrightarrow X\bar{X}$ (where $X$ denotes a standard model particle) can change 
the particle number, we can write
\begin{equation}
Y_\chi - Y_{\bar{\chi}} = C,\label{eq:Cdef}
\end{equation}
where $C$ is a strictly positive constant which defines the asymmetry between the particles
$\chi$ and antiparticles $\bar{\chi}$. We have assumed that $\chi$ is the majority component and $\bar{\chi}$ the minority component. Here, we are not particularly interested in 
the mechanism which generates the asymmetry, only that the asymmetry has been created well before particle freeze-out\footnote{If the asymmetry in the dark sector is linked to the baryonic asymmetry as discussed in the 
introduction, we would expect that $n_\chi - n_{\bar{\chi}} \approx n_b - n_{\bar{b}} 
\approx 6\times 10^{-10}n_\gamma$. Using the present entropy density $s \simeq 7.04 n_\gamma$~\cite{Olive} gives an estimate 
for the dark sector asymmetry, $C \sim  \mathcal{O}(10^{-11})$.}. 

The Boltzmann equations~\eqref{eq:dYasym1} can then be written
\begin{align}
\frac{dY_{\chi}}{dx}&=-A(x)\left(Y_\chi^2 - CY_{\chi} - P\right),\nonumber\\
\frac{dY_{\bar{\chi}}}{dx}&=
-A(x)\left(Y_{\bar{\chi}}^2 + CY_{\bar{\chi}} - P\right),\label{eq:dYasym}
\end{align}
where
\begin{equation}
P \equiv Y_\chi^{eq}Y_{\bar{\chi}}^{eq} = \left(\frac{0.145\,g_\chi}{g_{\star}}\right)^2x^3e^{-2x}.\label{eq:Pdef}
\end{equation}
The present day DM density is determined by solving the system~\eqref{eq:dYasym} in the limit $x\rightarrow \infty$ with the total density being the sum of the $\chi$ and $\bar{\chi}$ densities
\begin{align}
\Omega_{DM} h^2 &= \Omega_\chi h^2 + \Omega_{\bar{\chi}} h^2\nonumber\\
& = 2.74\times 10^8\, m_\chi\left[Y_\chi(\infty) + Y_{\bar{\chi}}(\infty)\right],\label{eq:omegadm}
\end{align}
where $Y(\infty) \equiv Y(x\rightarrow\infty)$. Note that setting $C = 0$ only establishes an equality between the particle and antiparticle 
number densities, i.e. $n_\chi = n_{\bar{\chi}}$. The particles  and anti-particles are
still distinct and they must be counted separately. For this reason 
the annihilation cross section needed to produce the observed DM abundance in 
asymmetric models with $C = 0$ is typically twice that of the symmetric case. 

Using~\eqref{eq:Cdef}, we also see that the introduction of an asymmetry places a lower 
bound on the relic density with $\Omega_{DM}h^2 > 2.74\times 10^8 m_\chi C$. This is a 
characteristic feature of asymmetric models in which the relic abundance is typically 
fixed by the asymmetry, $C$, as opposed to the symmetric case where the annihilation 
cross section, $\langle\sigma v\rangle$, determines the final density.  

In the coming sections we will see that, if DM particle freeze-out occurs during a 
braneworld expansion era, not only is the relic density enhanced with respect to the 
canonical result, but that the asymmetric DM behaves very much like symmetric DM in 
that the relic density is determined by the annihilation cross section and is independent 
of the asymmetry.

\subsection{Density evolution}
\label{sec:dens_ev}

\begin{figure}[tbp]
\centering % \begin{center}/\end{center} takes some additional vertical space
\includegraphics[scale=0.4,trim=15 125 40 180,clip=true]{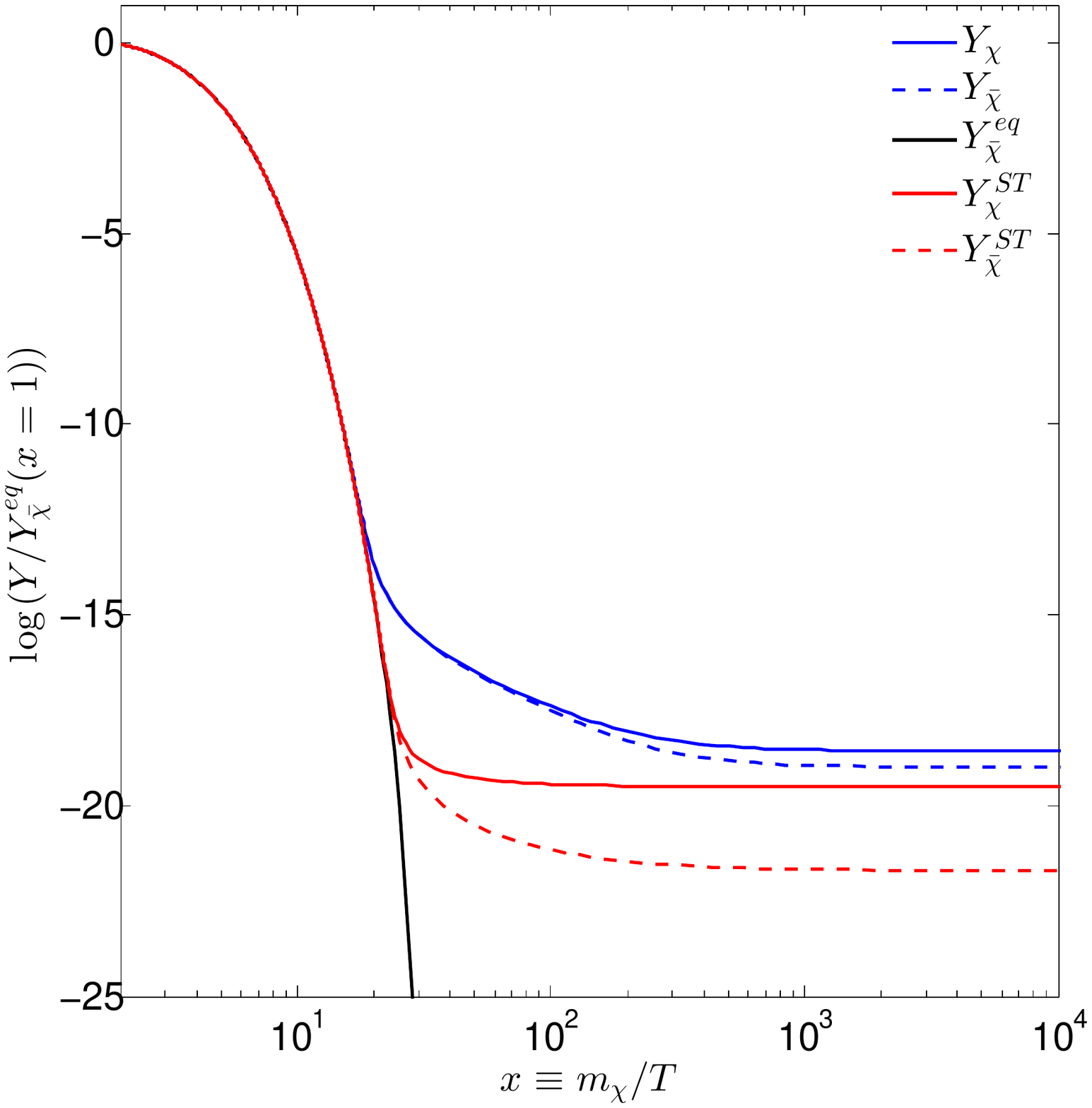}
\hfill
\includegraphics[scale=0.4,trim=15 125 40 180,clip=true]{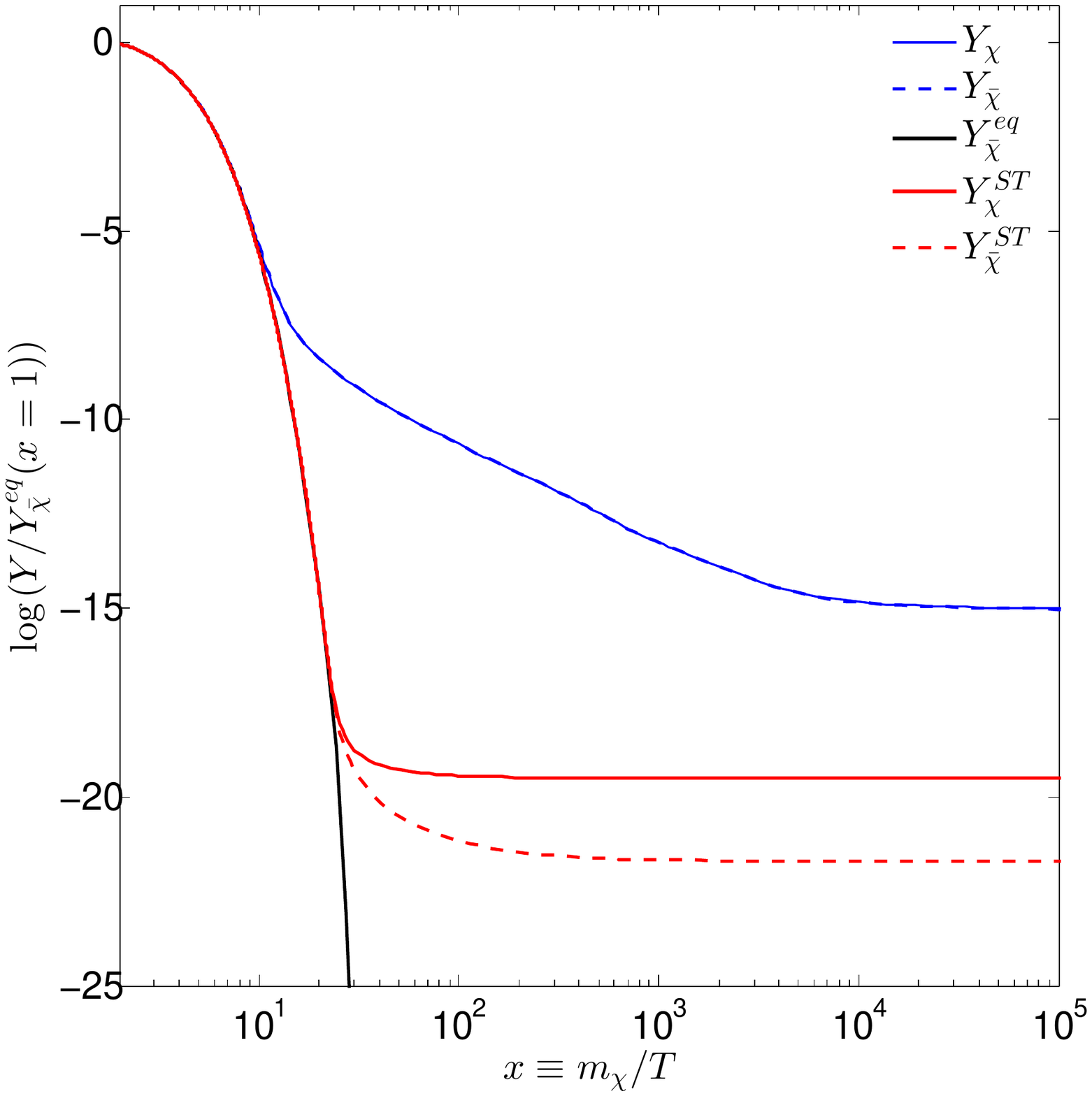}
\caption{\label{fig:Yevol} Evolution of the comoving number densities $Y_\chi$ (solid blue) 
and $Y_{\bar{\chi}}$ (dashed blue) as a function of $x$ in the RSII braneworld scenario 
with $M_5 = 10^6$ GeV (left panel) and $M_5 = 10^5$ GeV (right panel). The results have been calculated for an asymmetry $C = 4\times 10^{-12}$ and a WIMP with mass $m_\chi = 100$ GeV, 
$g_\chi = 2$ and annihilation cross section $\langle\sigma v \rangle = 
5\times 10^{-26}$ cm$^3$s$^{-1}$. Shown for reference are the comoving densities of the majority and minority components in the standard 
cosmology (red) as well as the equilibrium density $Y_{\bar{\chi}}^{eq}$ (black).}
\end{figure}

The evolution of the comoving number density in the braneworld scenario is determined
from the system of Boltzmann equations~\eqref{eq:dYasym} with the expansion rate $H(x)$ in
the factor $A(x)$ obtained from the modified Freidmann equation~\eqref{eq:brane_hub}.
Numerical solution of the equations was performed for an asymmetry $C = 4\times 10^{-12}$ 
and a WIMP with mass $m_\chi = 100$ GeV, $g_\chi = 2$ and annihilation cross section 
$\langle\sigma v\rangle = 5\times 10^{-26}$ cm$^{3}$s$^{-1}$. The results are shown in 
Figure~\ref{fig:Yevol}. We have considered the two cases; $M_5 = 10^6$ GeV and 
$M_5 = 10^5$ GeV (recall that smaller values of $M_5$ represent a greater departure 
from the standard expansion history). The density evolution in the standard scenario 
is also shown for reference.

Initially both the $\chi$ and $\bar{\chi}$ components are in equilibrium and 
$Y_{\chi,\bar{\chi}}$ track their equilibrium number densities $Y_{\chi,\bar{\chi}}^{eq}$. As the universe expands, the interaction rate drops below the expansion rate and the (anti)particles depart from equilibrium. The freeze-out point is $x_f \sim 20$ in the standard scenario. From Figure~\ref{fig:Yevol} we see that the modified expansion rate in the braneworld cosmology leads to earlier particle freeze-out 
($x_f\sim 13$ for $M_5 = 10^6$ GeV and $x_f\sim 10$ for $M_5 = 10^5$ GeV). 
There is also an enhanced relic density. 
A comparison of the left and right panels shows the effects are amplified for smaller values of 
$M_5$ with the relic density being boosted by a factor of $\sim 4$ and $\sim 160$ for 
$M_5 = 10^6$ GeV and $M_5 = 10^5$ GeV respectively. 

With the enhanced relic density, the asymmetry between the particles and antiparticles 
is essentially 'washed' out by the braneworld expansion era. Again this effect is most 
noticeable for smaller values of $M_5$ as can be seen in the right panel of 
Figure~\ref{fig:Yevol} where the number densities of the $\chi$ and $\bar{\chi}$ components coincide.

An interesting feature of Figure~\ref{fig:Yevol} is the evolution of the comoving 
number densities in the braneworld cosmology after particle freeze-out. The 
densities of the majority and minority components continue to decay even after decoupling 
and do not reach their asymptotic values until well after freeze-out. 
This behavior is characteristic of the RSII braneworld model where the asymptotic density 
is not reached until the standard expansion rate is restored~\cite{Okada1}, which for $M_5 = 10^5$ GeV and $M_5 = 10^6$ GeV occurs at $x_t \sim \mathcal{O}(10^3)$ and $x_t \sim \mathcal{O}(10^2)$ respectively\footnote{For the interval $x_f < x < x_t$ the comoving density in 
the braneworld cosmology behaves approximately as $Y_{\chi,\bar{\chi}} \sim 1/x$~\cite{Okada1}.}. 
This highlights the importance of maintaining the temperature dependence of $g_\star(T)$ 
in the numerical calculation.

\subsection{Analytical solution}
\label{sec:analsol}

Although the Boltzmann equations~\eqref{eq:dYasym} can not be solved exactly 
using analytical methods, we can find approximate solutions in limiting cases. 
For example, once the DM species has decoupled from the background thermal bath, 
the comoving number density $Y_{\chi(\bar{\chi})}\gg Y_{\chi(\bar{\chi})}^{eq}$ 
due to the exponential decay of the equilibrium density (see Fig.~\ref{fig:Yevol}). 
Hence the creation term proportional to $P$ in~\eqref{eq:dYasym} can be neglected for 
$x > x_f (\bar{x}_f)$ (where $x_f$ and $\bar{x}_f$ are the freeze-out points of the 
particle and antiparticle components respectively) and the reduced expressions 
can be integrated directly to yield~\cite{Iminniyaz1}
\begin{align}
Y_\chi(\infty) &= 
\frac{C}{1 - \exp{\left\{-C/Y_{(s)}(\infty)\right\}}},\label{eq:Ychi}\\
Y_{\bar{\chi}}(\infty) &= 
\frac{C}{\exp{\left\{C/Y_{(s)}(\infty)\right\}} - 1},\label{eq:Ychibar}
\end{align}
where
\begin{equation}
Y_{(s)}(\infty) \simeq \left[\int_{x_f}^{\infty}{A(x)\,dx}\right]^{-1}.\label{eq:ysymint}
\end{equation}
The subscript $(s)$ in $Y_{(s)}(\infty)$ is used to indicate that~\eqref{eq:ysymint} 
is the asymptotic 
solution for a symmetric DM species $\chi = \bar{\chi}$ \footnote{The Boltzmann 
equation for a symmetric DM species, $\chi = {\bar{\chi}}$, is given by,
\begin{equation}
\frac{dY}{dx} = -A(x)\left(Y^2 - Y_{eq}^2\right),\nonumber
\end{equation}
where $A(x)$ is given in~\eqref{eq:Adef} and $Y_{eq} \simeq 
0.145\left(g_\chi/g_\star\right) x^{3/2}e^{-x}$.}. 
To obtain~\eqref{eq:Ychi} and~\eqref{eq:Ychibar} we have assumed that the 
freeze-out points of the particles, $\chi$, and antiparticles, $\bar{\chi}$, 
are approximately equal, i.e. $x_f\simeq \bar{x}_f$, and thus the approximate 
solutions satisfy~\eqref{eq:Cdef}. 
We have also neglected the contribution from $1/Y(x_f)$ since we expect 
$Y(x_f) \gg Y(\infty)$~\cite{Kolb,Steigman}.

The density of the $\chi$ and $\bar{\chi}$ components depends sensitively on 
the ratio $C/Y_{(s)}(\infty)$. For large values of this ratio the contribution 
from the minority component ($Y_{\bar{\chi}}$) is exponentially suppressed and 
the density of the majority component approaches the asymmetry, 
$Y_\chi(\infty) \simeq C$. When this ratio is small, the asymmetry $C$ drops 
out of the expressions~\eqref{eq:Ychi} and \eqref{eq:Ychibar} and each component 
behaves like symmetric dark matter, 
$Y_\chi(\infty) \simeq Y_{\bar{\chi}}(\infty) \simeq Y_{(s)}(\infty)$. 
This second case is illustrated in the right panel of Figure~\ref{fig:Yevol} 
where the braneworld expansion rate has boosted the value of $Y_{(s)}(\infty)$ 
such that the $\chi$ and $\bar{\chi}$ densities are indistinguishable. 
The important quantity is the ratio $C/Y_{(s)}(\infty)$, so that even for 
large asymmetries $C$ the asymmetric DM can appear symmetric if the magnitude 
of $Y_{(s)}(\infty)$ is large enough.

Substituting~\eqref{eq:Ychi} and~\eqref{eq:Ychibar} into~\eqref{eq:omegadm} 
we obtain an approximate expression for the relic density $\Omega_{DM}h^2$,
\begin{equation}
\Omega_{DM}h^2 \simeq 2.74\times 10^8\,m_\chi\,C \coth{\left(\frac{C}{2 Y_{(s)}(\infty)}\right)},
\end{equation}
which in the two limiting cases discussed above becomes,
\begin{equation}
\Omega_{DM}h^2 \simeq 
\begin{cases}
\,2\times 2.74\times 10^8\,m_\chi\,Y_{(s)}(\infty), & \quad C/Y_{(s)}(\infty)\ll 1,\\
\,2.74\times 10^8\,m_\chi\,C, & \quad C/Y_{(s)}(\infty)\gg 1.\label{eq:omglim}
\end{cases}
\end{equation}

So far in this subsection we have not made any reference to the specific form of 
the integrand $A(x)$ in~\eqref{eq:ysymint}. This means that the arguments which have 
been developed can be applied to any cosmological scenario in which the relic 
density obeys~\eqref{eq:dYasym}. Indeed these effects have already been demonstrated 
for the quintessence~\cite{Gelmini,Iminniyaz2} and scalar-tensor~\cite{Gelmini} scenarios. 
Both of these models predict an enhanced expansion rate at the epoch of dark matter 
decoupling which was found to 'wash out' the asymmetry between the particles and antiparticles.

Focusing on the RSII braneworld cosmology, the integral~\eqref{eq:ysymint} has 
been computed in~\cite{Okada1} for the expansion rate~\eqref{eq:brane_hub} 
and in the limit $x_t \gg x_f$ becomes\footnote{To evaluate the 
integral~\eqref{eq:ysymint} the number of relativistic degrees of freedom 
$g_\star(T)$ has been fixed, i.e. $g_\star(T) = g_\star =$ const. We find that the most appropriate value of $g_\star(T)$ in the analytic expressions~\eqref{eq:ysymrs} and~\eqref{eq:ysymst} is given by $g_\star(T_t)$ and $g_\star(T_f)$ respectively, where $T_t = m_\chi/x_t$ is the transition temperature and $T_f = m_\chi/x_f$ is the freeze-out temperature.}
\begin{equation}
Y_{(s)}^{RS}(\infty) \simeq \frac{2.04\,x_t}{\sqrt{g_\star}\,m_\chi\,
M_{\mathrm{Pl}}\langle\sigma v \rangle}.\label{eq:ysymrs}
\end{equation} 
This expression is only valid if particle freeze-out occurs during the braneworld 
era $H \sim \rho$
. Thus, for smaller mass 
particles (which freeze out at lower temperatures), a later transition point (smaller $M_5$), is required for this approximation to hold. 
If DM decoupling occurs after the transition point, when the RSII expansion rate 
has reduced to the standard one, the canonical result is recovered,
\begin{equation}
Y_{(s)}^{ST}(\infty) \simeq \frac{3.79\,x_f^{ST}}{\sqrt{g_\star}\,m_\chi\, 
M_{\mathrm{Pl}}\langle\sigma v\rangle},\label{eq:ysymst}
\end{equation}
where $x_f^{ST}$ is the freeze-out point in the standard scenario. 
Inserting~\eqref{eq:ysymrs} into~\eqref{eq:Ychi} and~\eqref{eq:Ychibar} we see that, 
if either the annihilation cross section or the value of $M_5$ is large, we expect the 
$\bar{\chi}$ component to be exponentially suppressed, provided that $C/Y_{(s)}(\infty) \gtrsim \mathcal{O}(1)$, and the density of the majority 
component to be fixed by the asymmetry, $Y_\chi \simeq C$. If $\langle\sigma v\rangle$ 
and $M_5$ are small, such that $C/Y_{(s)}(\infty) \lesssim \mathcal{O}(1)$ (and 
$x_t \gg x_f^{ST}$), then the dark matter particles behave like symmetric dark matter and the relic density, using~\eqref{eq:omglim}, is given by,
\begin{equation}
\Omega^{RS}_{DM}h^2 \simeq 1.12\times 10^9\frac{x_t}{\sqrt{g_\star}
M_{\mathrm{Pl}}\langle\sigma v \rangle}.
\end{equation}
Comparing the relic density in the braneworld case with the standard cosmology result, 
we see that the abundance is boosted by a factor of~\cite{Okada1},
\begin{equation}
\frac{\Omega^{RS}}{\Omega^{ST}} \simeq 0.54\frac{x_t}{x_f^{ST}}.
\end{equation}
Therefore larger values of the transition point $x_t$ will result in greater 
enhancements of the DM relic abundance as we expected. To compensate, the annihilation 
cross section will have to be increased in order to provide the observed relic abundance~\eqref{eq:dm_abun}.

\section{Relic density constraints}
\label{sec:parbound}

\begin{figure}[tbp]
\centering % \begin{center}/\end{center} takes some additional vertical space
\includegraphics[scale=0.55,trim=50 175 75 210,clip=true]{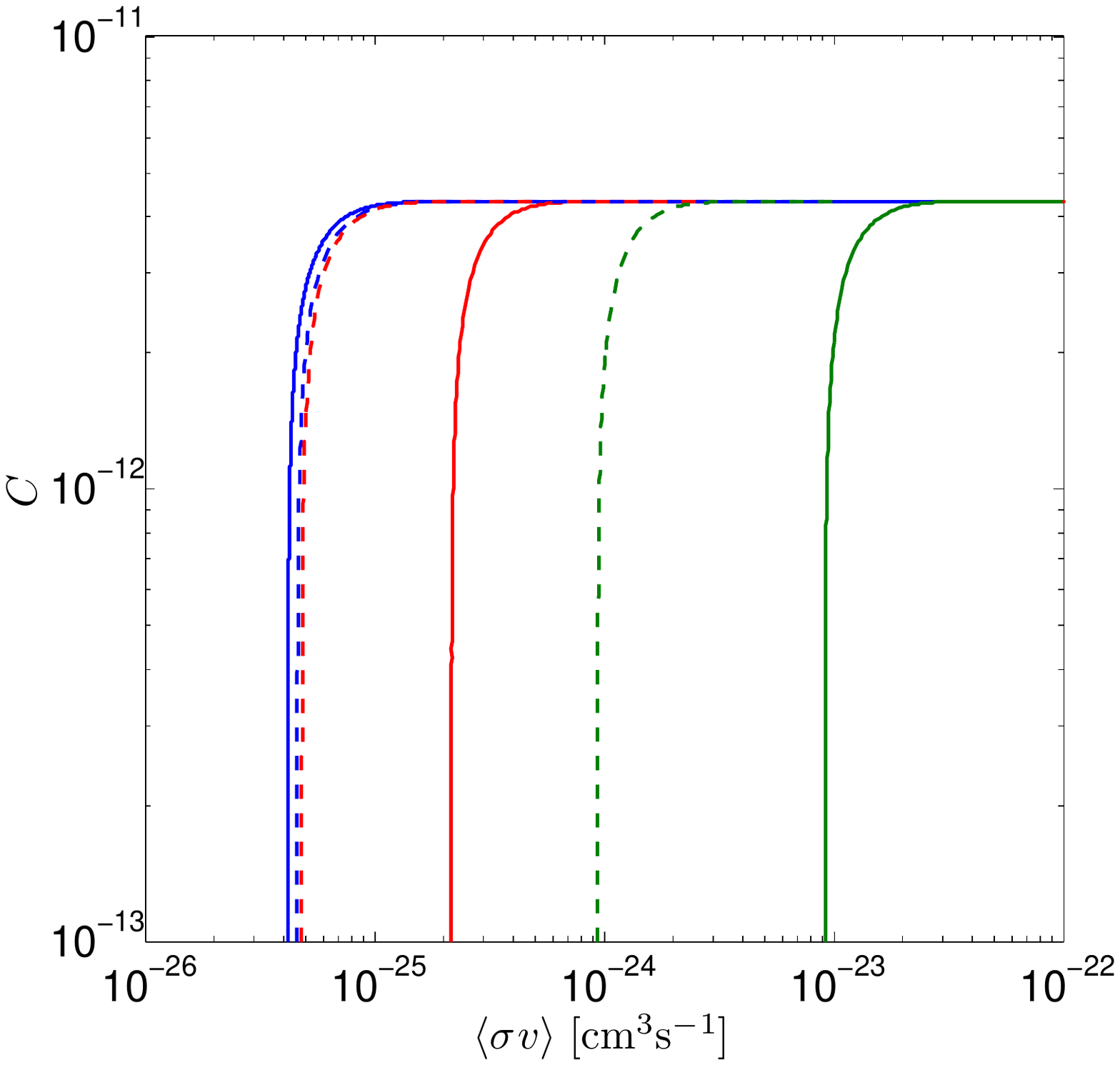}
\caption{\label{fig:sigCvarM5} Contours in the $(\langle\sigma v\rangle,m_{100}C)$ 
plane which satisfy the relic abundance bound $\Omega_{DM}h^2 = 0.1187$~\cite{Lahav} 
for varying $M_5$. The solid blue, red and green curves represent the standard, 
$M_5 = 10^6$ GeV and $M_5 = 10^5$ GeV scenarios respectively for a WIMP with mass $m_\chi = 100$ GeV. The dashed curves give the corresponding 
scenarios with $m_\chi = 10$ GeV.}
\end{figure}

The magnitude of the five dimensional Planck mass, $M_5$, and the asymmetry, $C$, can 
be constrained by the observed DM relic density~\eqref{eq:dm_abun}. 
In Figure~\ref{fig:sigCvarM5} we plot the contours in the $(\langle\sigma v\rangle, m_{100}C)$ plane which give the correct relic abundance for varying $M_5$. 
The red and green curves correspond to $M_5 = 10^6$ GeV and $M_5 = 10^5$ GeV respectively and the blue curve gives the standard cosmology result. 
We have considered both $m_\chi = 100$ GeV (solid) and $m_\chi = 10$ GeV (dashed).  

Two distinct regions are apparent. In the first, the curves are vertical and the relic 
abundance is independent of the asymmetry. In this region the ratio $C/Y_{(s)}(\infty)$ 
is small and the $\chi$ and $\bar{\chi}$ components behave like symmetric DM where the density is determined by the annihilation cross section, see~\eqref{eq:omglim}, \eqref{eq:ysymrs} and\eqref{eq:ysymst}. 
As the asymmetry increases so does the magnitude of the ratio $C/Y_{(s)}(\infty)$ and the contours 
transition into a region which is strongly asymmetric, i.e. the relic density is 
now fixed by the asymmetry $C$ and is independent of the annihilation cross section. 
In this regime the density of the minority component is exponentially suppressed 
and the DM abundance is given by $\Omega_{DM}h^2 \sim m_\chi Y_\chi \sim m_\chi C$. 
For the observed density~\eqref{eq:dm_abun} this occurs at $m_{100}C = 4.33\times 10^{-12}$ 
\footnote{There is also an intermediate region where the abundance depends on both 
the asymmetry and the annihilation cross section. Here the minority component 
freezes out shortly after the majority component and their final densities are comparable. For a discussion on each regime and their relation to 
the freeze-out of the $\chi$ and $\bar{\chi}$ components see~\cite{Gelmini}.}.

Each of the iso-abundance contours is accurately described by the relation,
\begin{equation}
\langle\sigma v\rangle \sim \frac{1}{C}\;
\coth^{-1}{\left(\frac{\omega}{C}\right)},\label{eq:siginC}
\end{equation}
where $\omega = \Omega_{DM}h^2/(2.74\times 10^8\,m_\chi)$\footnote{Again, this will be true of any cosmology in which the symmetric solution $Y_{(s)}(\infty) \sim 1/\langle\sigma v\rangle$.}. 
The appropriate form of~\eqref{eq:siginC} depends on whether we are in the region 
$x_f < x_t$ or $x_f > x_t$, in which case we would use the approximate 
formulas~\eqref{eq:ysymrs} or~\eqref{eq:ysymst} respectively. In the first case, 
freeze-out occurs prior to the transition point and the annihilation cross section 
is shifted to larger values as the magnitude of $M_5$ decreases (or as $x_t$ increases). 
The shift is more pronounced for the $m_\chi = 100$ GeV case since these particles 
freeze-out earlier compared to the $m_\chi = 10$ GeV particles (see the discussion 
in section~\ref{sec:brane}).

\begin{figure}[tbp]
\centering % \begin{center}/\end{center} takes some additional vertical space
\includegraphics[scale=0.4,trim=15 125 40 180,clip=true]{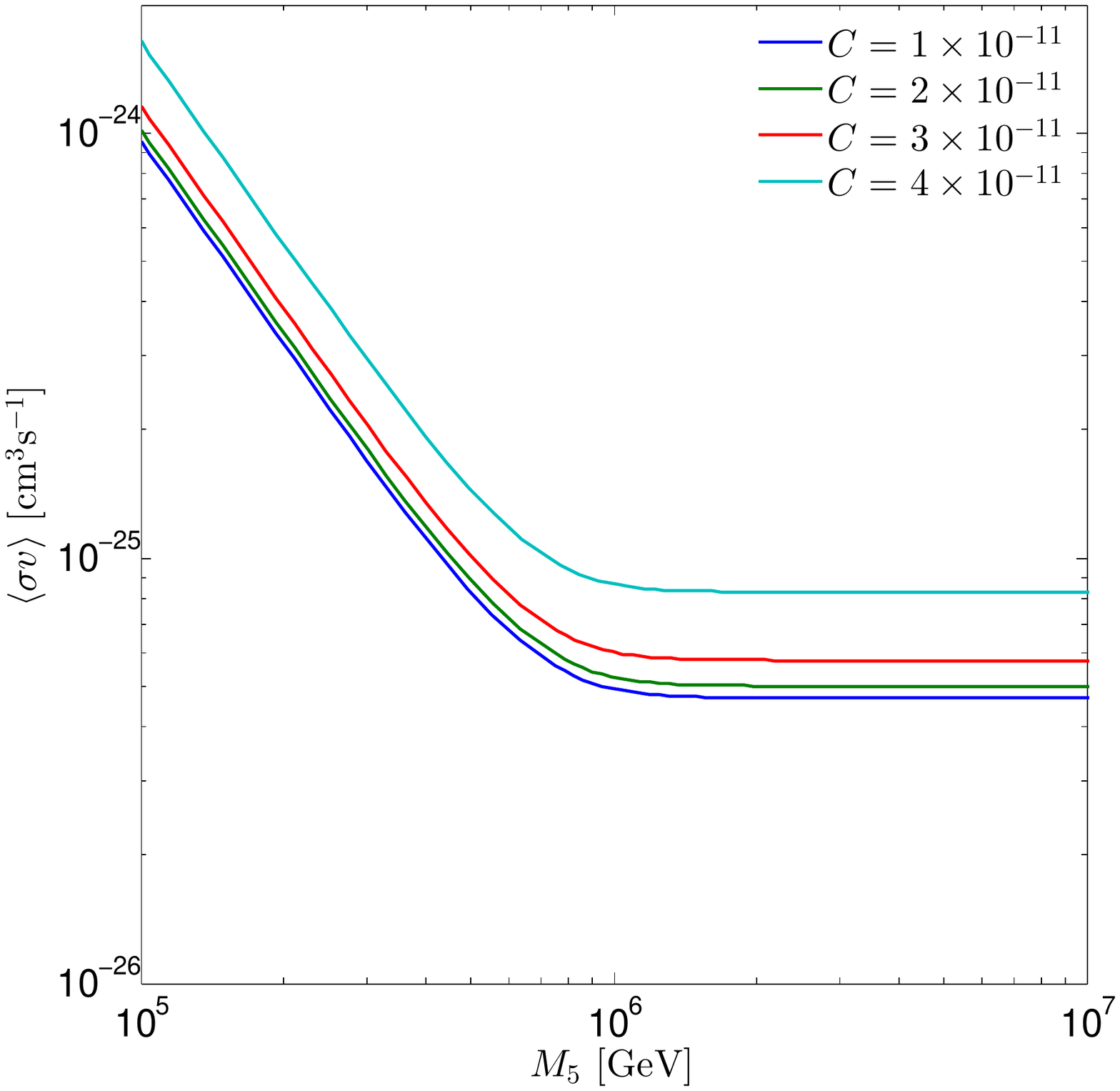}
\hfill
\includegraphics[scale=0.4,trim=15 125 40 180,clip=true]{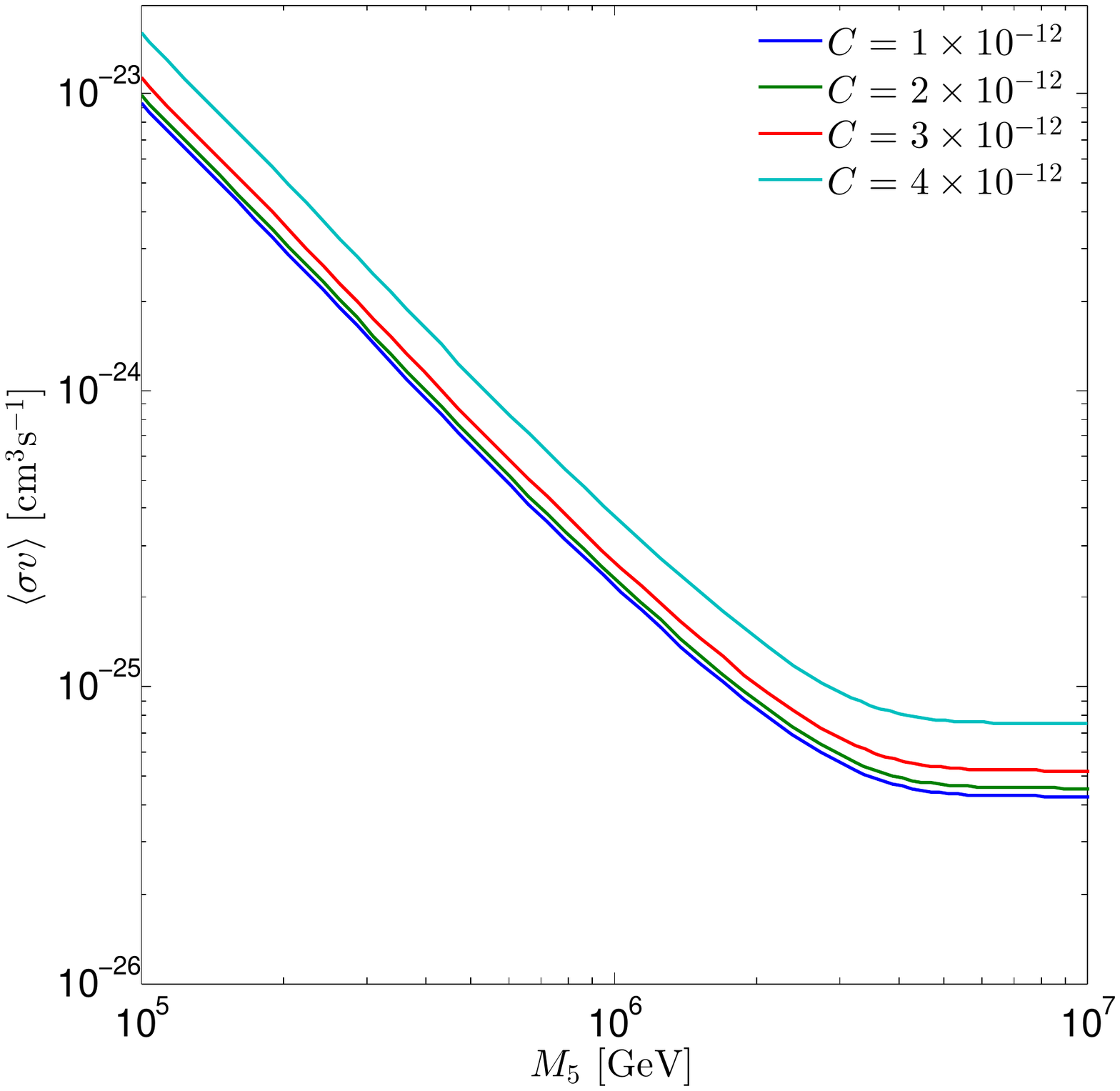}
\caption{\label{fig:M5sigvarC} The required annihilation cross section in the 
braneworld scenario as a function of the five dimensional Planck mass $M_5$. 
The results have been calculated for varying asymmetries with 
$m_\chi = 10$ GeV (left panel) and $m_\chi = 100$ GeV (right panel).}
\end{figure}

Using the observed relic density~\eqref{eq:dm_abun} we can also determine the 
required annihilation cross section as a function of the five dimensional Planck 
mass $M_5$. These results are shown in Figure~\ref{fig:M5sigvarC} for different values 
of the asymmetry $C$. The left panel corresponds to a WIMP with mass $m_\chi = 10$ GeV 
and the right panel to $m_\chi = 100$ GeV. 
In each plot we see that the required annihilation cross section is increased by up to 
several orders of magnitude compared to the standard result 
(see the blue curves in Figure~\ref{fig:sigCvarM5}).
From the earlier discussion (see section~\ref{sec:brane}) we recall that the 
annihilation cross section is only enhanced if freeze-out occurs during a braneworld type 
expansion era, i.e. for $x_f < x_t$. In the standard cosmology freeze-out occurs at 
approximately $x_f^{ST}\sim 20$ so we require $x_t \gtrsim 20$. Using the 
definition of $x_t$,~\eqref{eq:xtdef}, this corresponds to 
$M_5 \lesssim 9\times 10^5$ GeV and $M_5 \lesssim 5\times 10^6$ GeV for 
$m_\chi = 10$ GeV and $m_\chi = 100$ GeV respectively. 
In this region the curves approximately obey  
\begin{equation}
\langle\sigma v\rangle \propto \frac{m_\chi}{\Omega_{DM}h^2}M_5^{-3/2}.
\end{equation}
For larger values of $M_5$ (smaller $x_t$) the standard expansion law is recovered 
prior to DM decoupling and the curves saturate to the standard result.

\section{Annihilation rate for asymmetric DM}
\label{sec:annrate}
The modified expansion rate in the braneworld scenario requires an enhanced 
annihilation cross section in order to provide the observed DM density. In this section 
we discuss the impact this result has on the annihilation rate of asymmetric DM, $\Gamma_{(a)}$. 
We find that although the density of the minority component is exponentially suppressed 
in asymmetric DM models, the increased annihilation cross section may compensate for this 
and produce an amplified detection signal. This effect 
has been pointed out by~\cite{Gelmini} for the kination and scalar-tensor scenarios 
and is in contrast to the usual expectation that the asymmetric signal is reduced 
with respect to the symmetric one. 

To compare the expected detection signals in the two cases we first write the 
annihilation rate of symmetric dark matter in the standard scenario~\cite{Gelmini},
\begin{equation}
\Gamma_{(s)}^{ST} = \frac{1}{2}\langle\sigma_s v\rangle^{ST}
\left(\frac{\rho_{DM}}{m_\chi}\right)^2,
\end{equation}
where $\langle\sigma_s v\rangle^{ST}$ is the self annihilation cross section of the 
self-conjugate particles $\chi = \bar{\chi}$ in the standard cosmological scenario, 
$\rho_{DM}$ is the DM energy density, $\rho_{DM} = \Omega_{DM}\rho_c$ 
(where $\rho_c$ is the critical density), and the factor of $1/2$ is necessary since the annihilating particles are identical. The required annihilation 
cross section in the standard scenario is found numerically to be 
$\langle\sigma_s v\rangle^{ST} = 2.03\times 10^{-26}$ cm$^3$s$^{-1}$ for 
$m_\chi = 100$ GeV and $\langle\sigma_s v\rangle^{ST} =2.21\times 10^{-26} $ cm$^3$s$^{-1}$ 
for $m_\chi = 10$ GeV.

\begin{figure}[tbp]
\centering % \begin{center}/\end{center} takes some additional vertical space
\includegraphics[scale=0.7,trim=100 250 100 200,clip=true]{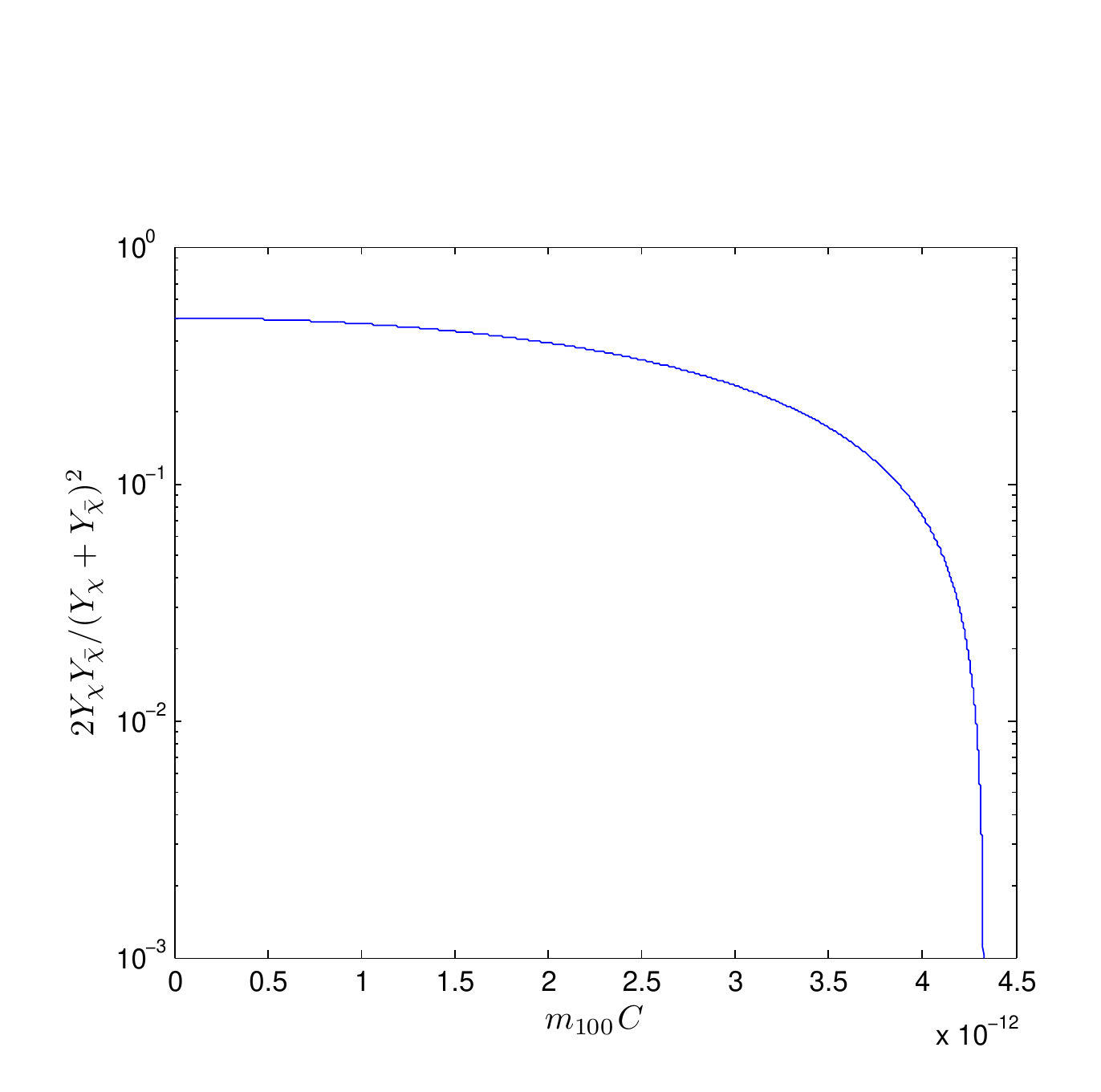}
\caption{\label{fig:RatiovC} The asymmetric damping factor defined 
in~\eqref{eq:dampfac} as a function of $m_{100}C$ where $m_{100}$ is the DM mass 
in units of 100 GeV and $C$ is the DM asymmetry.}
\end{figure}

The annihilation rate for asymmetric dark matter in the braneworld scenario can be written, 
using~\eqref{eq:Cdef} and~\eqref{eq:omegadm}, as 
\begin{equation}
\Gamma_{(a)} = \langle \sigma v \rangle \frac{\rho_{\chi}\rho_{\bar{\chi}}}{m_{\chi}^{2}}
= \langle\sigma v\rangle \left(\frac{\rho_{DM}}{m_\chi}\right)^2
\frac{Y_\chi Y_{\bar{\chi}}}{(Y_\chi + Y_{\bar{\chi}})^2},
\end{equation}
and the ratio of the two annihilation rates is then
\begin{equation}
\frac{\Gamma_{(a)}}{\Gamma_{(s)}^{ST}} = \frac{\langle\sigma v\rangle}
{\langle\sigma_s v\rangle^{ST}}\frac{2Y_\chi Y_{\bar{\chi}}}{(Y_\chi + 
Y_{\bar{\chi}})^2}.\label{eq:gamrat}
\end{equation}
Once the relic density $\Omega_{DM}h^2$ has been specified, the damping factor 
$\gamma\equiv 2Y_{\chi}Y_{\bar{\chi}}/(Y_{\chi} + Y_{\bar{\chi}})^2$ can be expressed 
explicitly in terms of the asymmetry $C$;
\begin{equation}
\gamma\equiv \frac{2Y_{\chi}Y_{\bar{\chi}}}{(Y_\chi + Y_{\bar{\chi}})^2} = 
\frac{\omega^2 - C^2}{2\omega^2}\label{eq:dampfac},
\end{equation}
where $\omega = \Omega_{DM}h^2/(2.74\times 10^8\,m_\chi)$. The magnitude of this 
factor varies from $\gamma\simeq 1/2$ for small $C$ to $\gamma = 0$ at $C = \omega$. 
This relationship has been plotted in Figure~\ref{fig:RatiovC} for the observed 
DM density~\eqref{eq:dm_abun} where we see that, for $m_{100}C\lesssim 4\times 10^{-12}$, 
the damping factor $\gamma \sim \mathcal{O}(10^{-1})$ and then rapidly drops to zero thereafter. 
Given that the required annihilation cross section can be several orders of 
magnitude larger in the braneworld scenario compared to the symmetric one in the 
standard cosmology (see Figure~\ref{fig:M5sigvarC}), the ratio of the annihilation 
cross sections in~\eqref{eq:gamrat} can easily compensate for the asymmetric 
damping factor for a large range of asymmetries, $C$. 

\begin{figure}[tbp]
\centering % \begin{center}/\end{center} takes some additional vertical space
\includegraphics[scale=0.4,trim=15 125 40 180,clip=true]{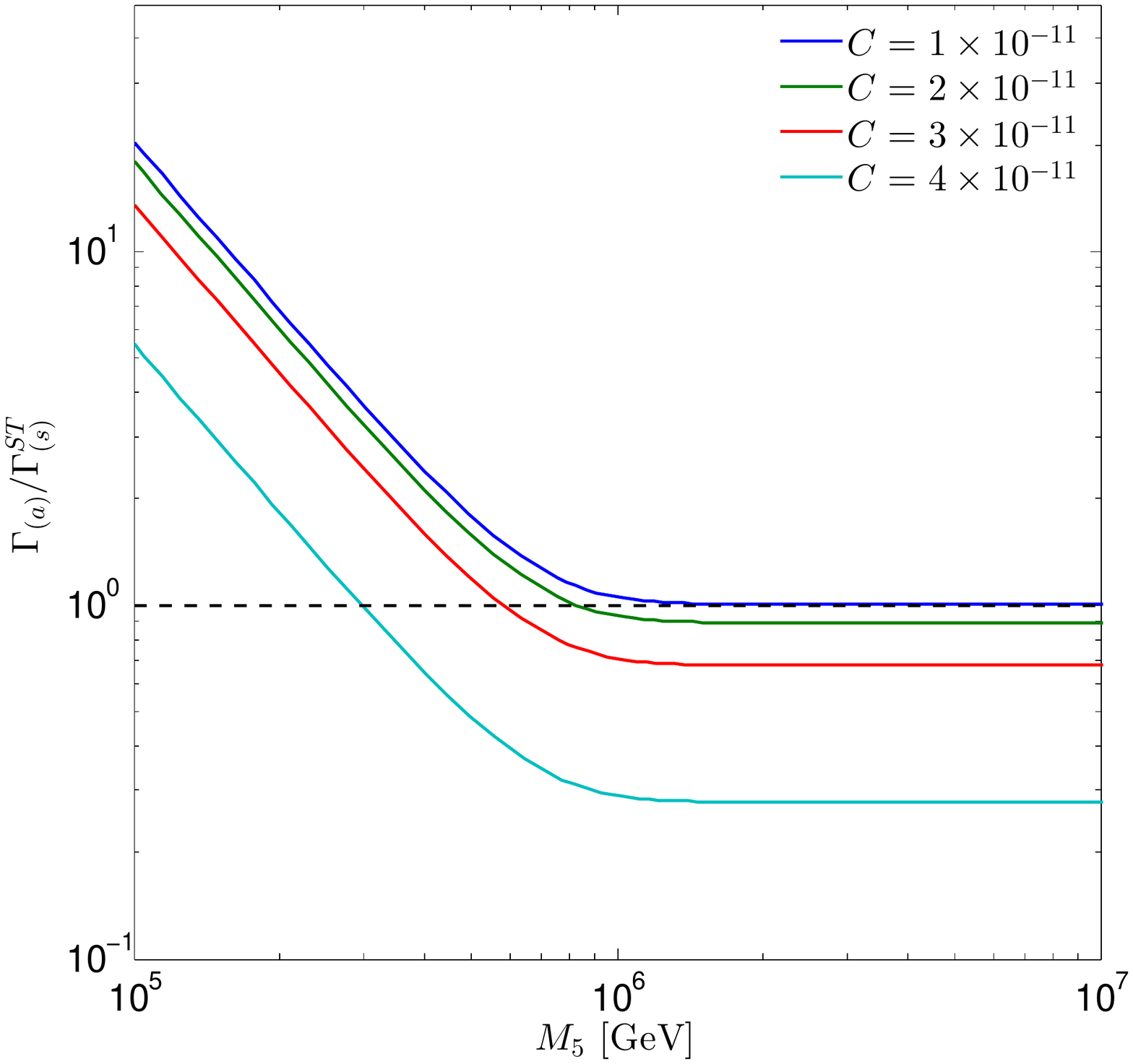}
\hfill
\includegraphics[scale=0.4,trim=15 125 40 180,clip=true]{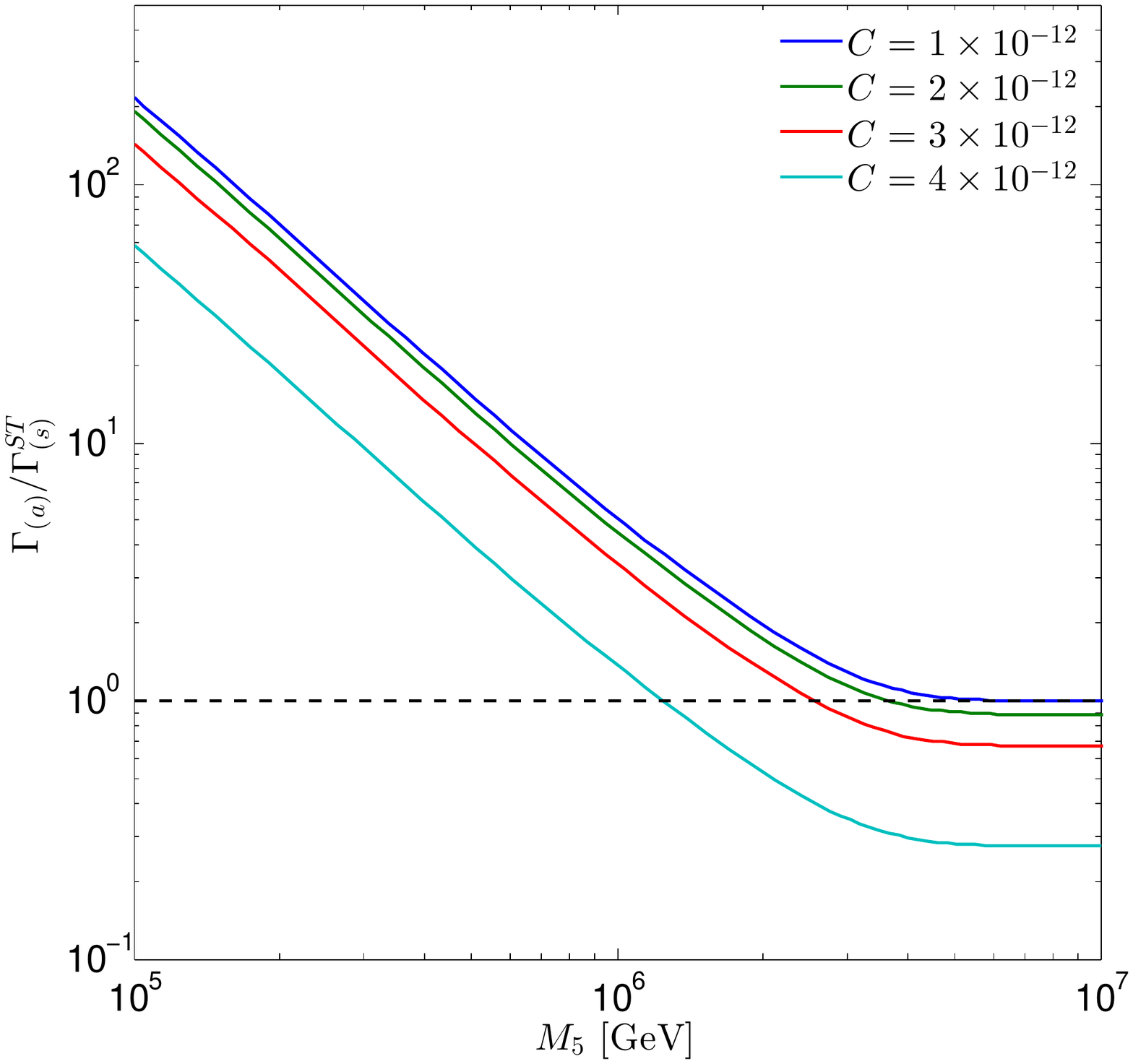}
\caption{\label{fig:M5ratio} Ratio of the asymmetric annihilation rate in 
the braneworld cosmology to the symmetric annihilation rate in the standard scenario 
as a function of $M_5$ with $m_\chi = 10$ GeV (left panel) and $m_\chi = 100$ GeV (right panel). 
The horizontal dashed line corresponds to $\Gamma_{(a)} = \Gamma_{(s)}^{ST}$.}
\end{figure}

The ratio $\Gamma_{(a)}/\Gamma_{(s)}^{ST}$, given by~\eqref{eq:gamrat}, can be evaluated 
using the cross sections calculated in the previous section. This ratio is shown 
in Figure~\ref{fig:M5ratio} as a function of $M_5$ for different values of the asymmetry $C$. 
As predicted, the asymmetric annihilation rate in the 
braneworld scenario can exceed the symmetric one in the standard scenario for a 
significant region of the parameter space. In particular we find that, for 
$m_{100}C = 10^{-12}$ and $M_5 = 10^5$ GeV, the asymmetric signal is $\sim 200(20)$ 
times larger than the symmetric signal in the standard cosmology for 
$m_\chi = 100(10)$ GeV. Even larger asymmetries such as 
$m_{100}C = 4\times 10^{-12}$ can produce a signal enhanced by a factor of 
$\sim 60(6)$ for $m_\chi = 100(10)$ GeV.

Although enhanced signals for asymmetric DM are possible in the braneworld 
scenario for certain parameter combinations, it is important to check that such 
combinations satisfy the observational constraints on the DM annihilation cross 
section. In the next section we review our results in light of the gamma ray 
data from the Fermi space telescope.

\section{Fermi-LAT constraints}
\label{sec:Fermi}
\begin{figure}[tbp]
\centering % \begin{center}/\end{center} takes some additional vertical space
\includegraphics[scale=0.45,trim=50 175 75 210,clip=true]{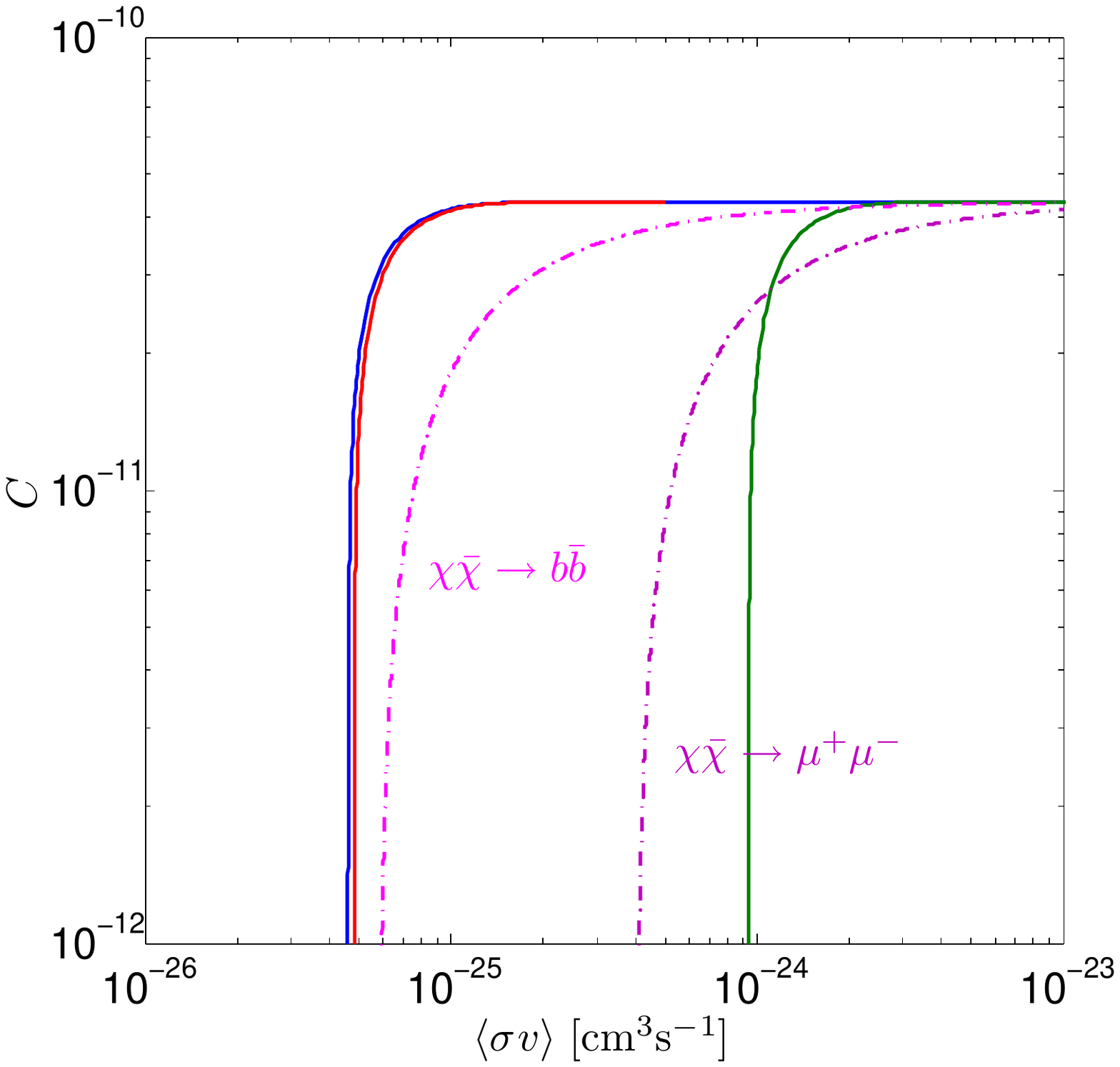}
\hfill
\includegraphics[scale=0.45,trim=50 175 75 210,clip=true]{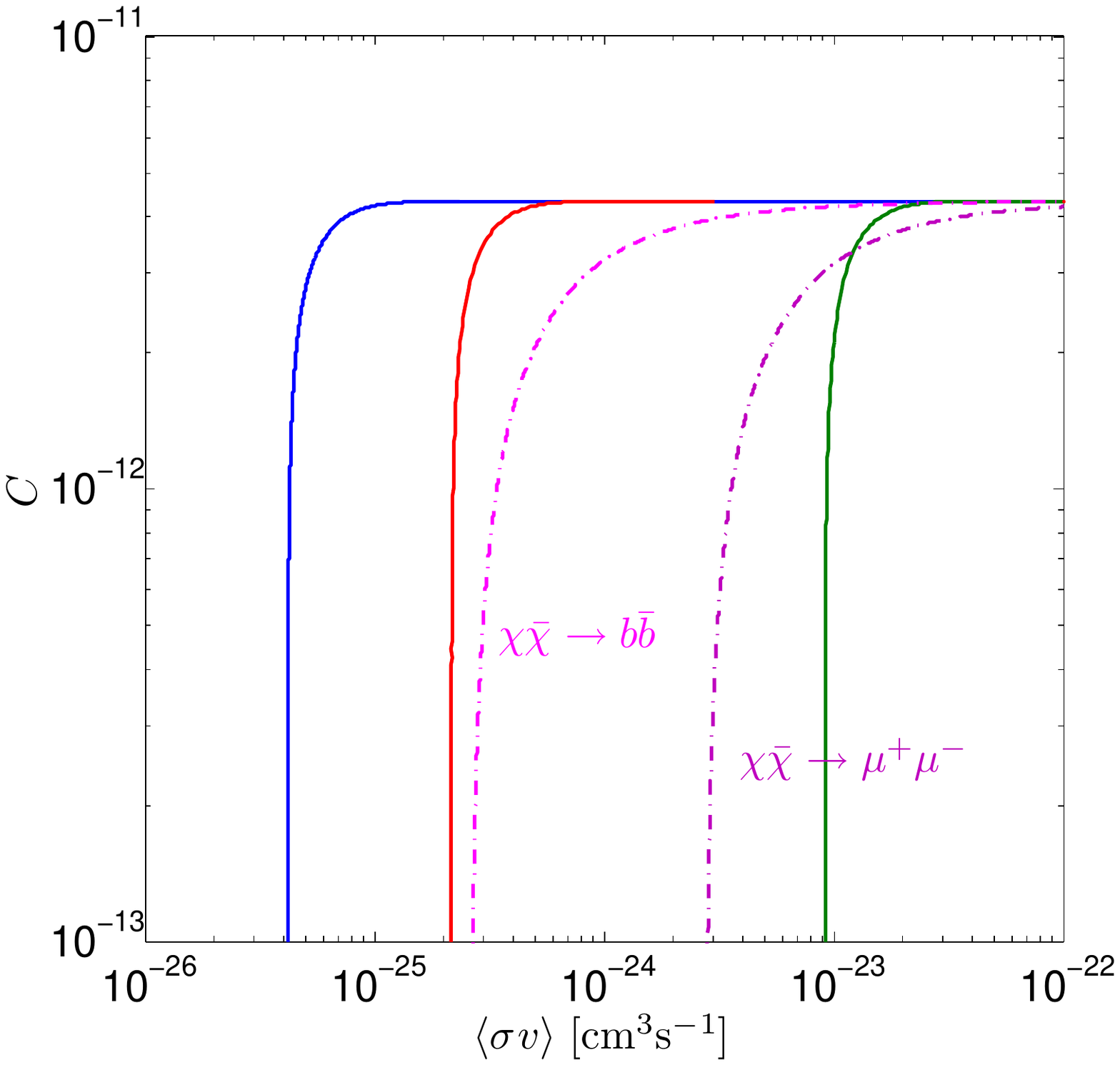}
\caption{\label{fig:Fermibound} Same as Figure~\ref{fig:sigCvarM5} with Fermi-LAT 
bounds~\cite{Fermi} (dashed-pink) superimposed for $m_\chi = 10$ GeV (left panel) 
and $m_\chi = 100$ GeV (right panel).}
\end{figure}

In addition to the relic density observation, data is available from the 
Fermi Large Area Telescope (Fermi-LAT) which searches for gamma rays from 
satellite galaxies of the Milky Way, such as those originating from DM 
annihilations~\cite{Fermi}. Depending on the annihilation channel, the gamma ray 
data can be used to place upper bounds on the DM annihilation cross section assuming 
the DM species constitutes all of the DM density. Therefore, the predicted asymmetric detection signal will satisfy the Fermi bound if,
\begin{equation}
\frac{\Gamma_{(a)}}{\Gamma_{\mathrm{Fermi}}} = \frac{\langle\sigma v\rangle}
{\langle\sigma_s v\rangle_{\mathrm{Fermi}}}\frac{2Y_\chi Y_{\bar{\chi}}}{\left(Y_\chi^2 + Y_{\bar{\chi}}^2\right)} < 1,
\end{equation}
where $\langle\sigma_s v\rangle_{\mathrm{Fermi}}$ is the inferred cross section bound 
from the Fermi gamma ray data. 
In terms of the asymmetric annihilation cross section, $\langle\sigma v\rangle$, 
and asymmetry, $C$, this condition can be expressed as (using~\eqref{eq:gamrat}),
\begin{equation}
C >\omega\left(1 - 2\frac{\langle\sigma_s v\rangle_{\mathrm{Fermi}}}
{\langle\sigma v\rangle}\right)^{1/2}.
\end{equation}
In Figure~\ref{fig:Fermibound} we superimpose these constraints on the iso-abundance 
contours of Figure~\ref{fig:sigCvarM5} (the constraints are shown as the pink dashed 
curves with the area beneath the curves excluded for that particular annihilation channel). 
For illustration we have only displayed the bounds for the 
$\chi\bar{\chi}\rightarrow b\bar{b}$ and 
$\chi\bar{\chi}\rightarrow \mu^+\mu^-$ channels which, for a WIMP with mass 
$m_\chi = 100$ GeV, are $\langle\sigma v\rangle_{\mathrm{Fermi}} = 1.31\times 10^{-25}$ 
cm$^3$s$^{-1}$ and $\langle\sigma v\rangle_{\mathrm{Fermi}} = 1.38\times 10^{-24}$ 
cm$^3$s$^{-1}$ respectively. 
For the $m_\chi = 10$ GeV case the bounds are more stringent with 
$\langle\sigma v\rangle_{\mathrm{Fermi}} = 2.90\times 10^{-26}$ cm$^3$s$^{-1}$ and 
$\langle\sigma v\rangle_{\mathrm{Fermi}} = 2.01\times 10^{-25}$ cm$^3$s$^{-1}$ 
for the respective channels.

\begin{figure}[tbp]
\centering % \begin{center}/\end{center} takes some additional vertical space
\includegraphics[scale=0.45,trim=50 175 75 210,clip=true]{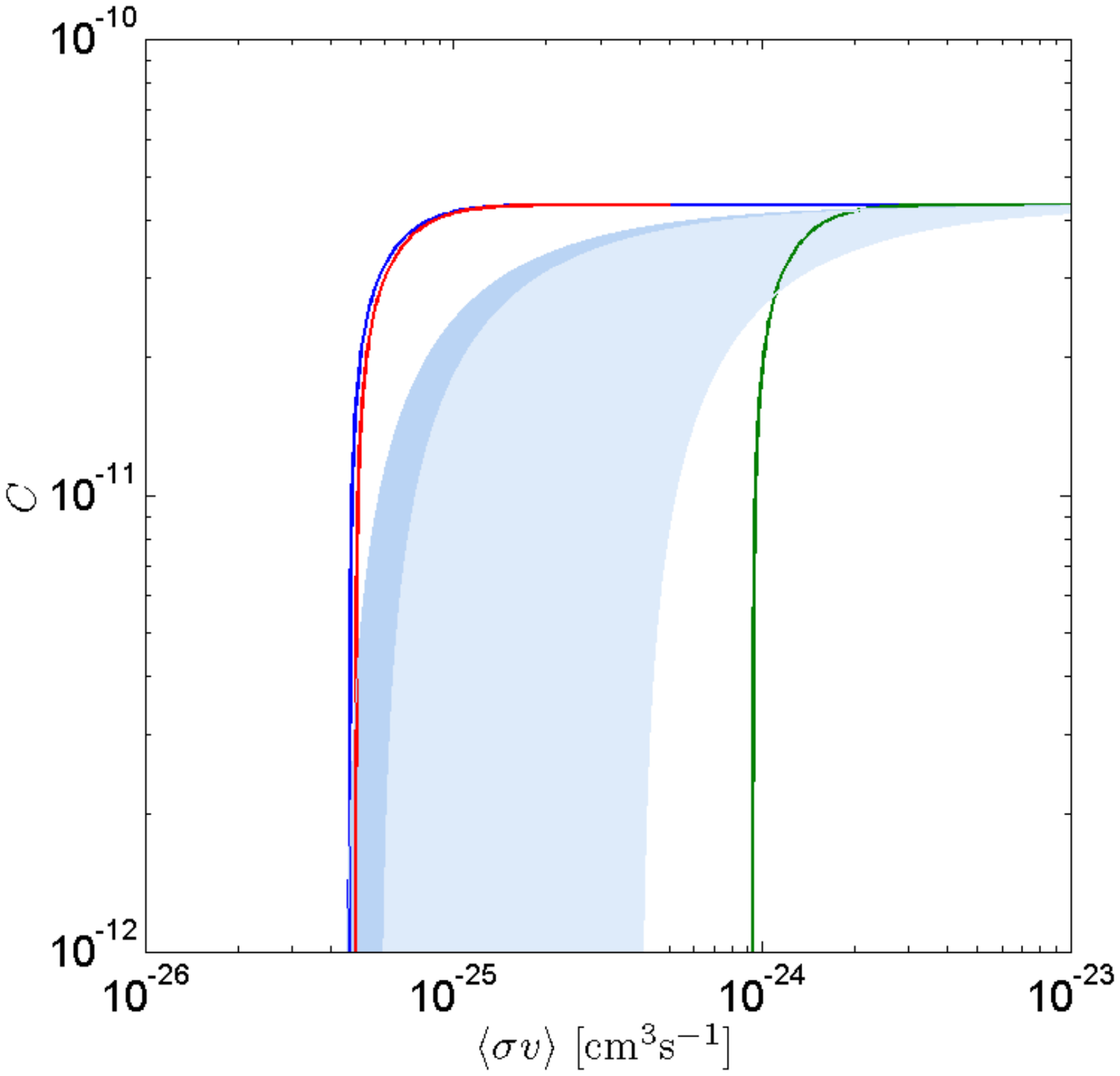}
\hfill
\includegraphics[scale=0.45,trim=50 175 75 210,clip=true]{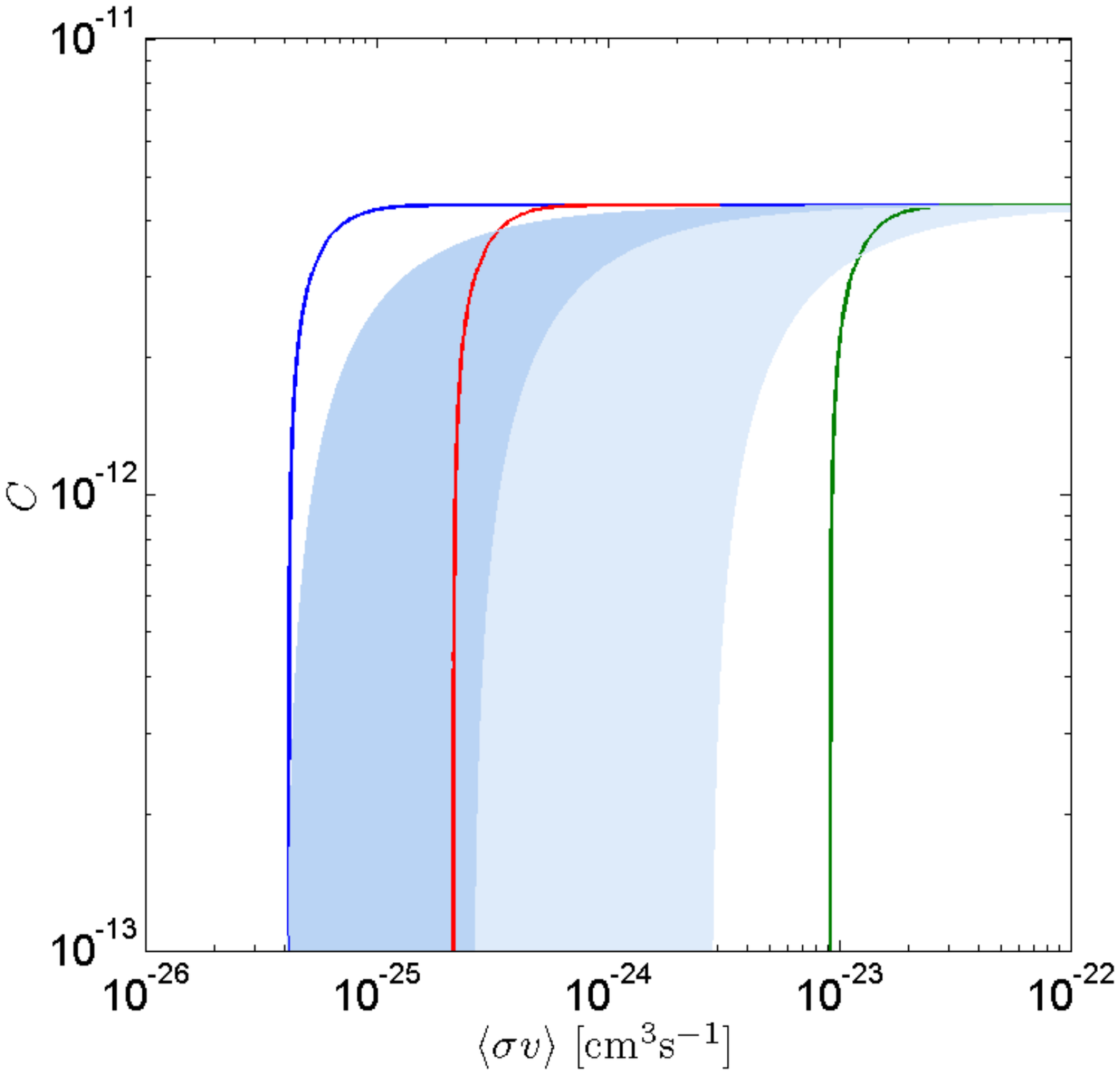}
\caption{\label{fig:enh_reg} The allowed regions in the $(\langle\sigma v\rangle,C)$ 
plane which give an enhanced annihilation cross section for $m_\chi = 10$ GeV (left panel) 
and $m_\chi = 100$ GeV (right panel). The darker shaded area corresponds to the 
region which satisfies the $\chi{\bar{\chi}}\rightarrow b\bar{b}$ bound and the combination of the dark and lighter shaded regions satisfy the 
$\chi\bar{\chi}\rightarrow \mu^+\mu^-$ bound~\cite{Fermi}.}
\end{figure}

The results in Figure~\ref{fig:M5ratio} should be compared with the constraints 
displayed in Figure~\ref{fig:Fermibound} in order to determine which combinations 
of $M_5$ and $C$ simultaneously satisfy the Fermi bounds whilst predicting an 
enhanced detection signal, $\Gamma_{(a)} > \Gamma_{(s)}^{ST}$. For example, for the 
combination $m_\chi = 100$ GeV and $M_5 = 10^6$ GeV, we see from the right 
panel of Figure~\ref{fig:Fermibound} that all values of the asymmetry $C$ are 
permitted by the Fermi-LAT data. Then, from the right panel of Figure~\ref{fig:M5ratio} 
we can see that for this value of $M_5$, the asymmetric annihilation rate is 
amplified, $\Gamma_{(a)} > \Gamma_{(s)}^{ST}$, for asymmetries up to $C\sim 4\times 10^{-12}$. 
In general, it is possible to have an enhanced annihilation rate which is consistent 
with the Fermi-LAT data, if the asymmetric annihilation cross section, 
$\langle\sigma v\rangle$, and asymmetry $C$, satisfy
\begin{equation}
\langle\sigma_s v\rangle^{ST} < \gamma\langle\sigma v\rangle
 < \langle\sigma_s v\rangle_{\mathrm{Fermi}},
\end{equation}
where $\gamma$ is the damping factor defined in~\eqref{eq:dampfac}. This region is plotted in Fig.~\ref{fig:enh_reg} and depends upon the appropriate 
annihilation channel.

\section{Conclusions}
\label{sec:concl}

In this article we have investigated the relic density and observational signatures of asymmetric dark matter models in the braneworld cosmology.
We have found that the decoupling of asymmetric dark matter and the evolution of the particle/antiparticle number densities is modified in the RSII braneworld scenario. 
In particular, the enhanced braneworld expansion rate leads to earlier particle freeze-out and an enhanced relic density. Additionally, the asymmetry between the DM particles and antiparticles is 'washed' out by the modified expansion rate and the relic abundance is determined by the 
annihilation cross section $\langle\sigma v \rangle$. 
In this respect the nominally asymmetric model behaves like symmetric dark matter. This effect was demonstrated analytically in section~\ref{sec:analsol} where we derived approximate expressions for the densities of the majority and minority DM components in the braneworld scenario and is consistent with other investigations of asymmetric dark matter in non-standard cosmologies which predict an enhanced expansion rate during the epoch of dark matter decoupling~\cite{Gelmini,Iminniyaz2}. 

Importantly, we have also accounted for the variation in the number of relativistic degrees of freedom as a function of temperature, $g_\star(T)$, in our numerical solution of the Boltzmann equation. This variation is particularly pertinent for the RSII braneworld model in which asymptotic behaviour of the number densities is not achieved until well after decoupling. Fixing $g_\star(T) =$ const. in the numerical integration can result in relic density predictions which are out by a factor of $\sim 2$ depending on the magnitude of $M_5$. 

Finally, in order to provide the observed relic density~\eqref{eq:dm_abun} the 
annihilation cross section must be boosted to compensate for the enhanced expansion rate. 
In this instance we find that the annihilation rate, 
$\Gamma_{\chi(\bar{\chi})}\propto \langle\sigma v\rangle$, of asymmetric DM in the 
braneworld cosmology is larger than the symmetric signal in the standard scenario 
despite the suppressed abundance of the minority DM component. This effect is contrary to the 
usual expectation which presumes a weaker asymmetric detection signal. We have also shown that 
it is possible to produce an amplified asymmetric detection signal which satisfies the 
observational constraints from Fermi-LAT~\cite{Fermi}.

Additional data from Fermi-LAT and other indirect detection probes~\cite{AMS,PAMELA} may further constrain the dark matter particle properties and in the process shed light on the cosmological conditions prior to BBN. 

Very recently new information about the very early universe has been provided by the BICEP2 experiment~\cite{BICEP2} which detected $B$-mode polarization of the CMB, indicating a large tensor-to-scalar ratio $r=0.20^{+0.07}_{-0.05}$. These primordial tensor perturbations are strong evidence of gravitational waves and are a generic prediction of inflationary cosmological models with large energy density~ \cite{Ellis2014a,Ellis2014b}. Whereas the BICEP2 result of $n_{s} \simeq 0.96$  for the infrared tilt of the scalar perturbations $n_{s}$ is in agreement with Planck~\cite{Planck22} and WMAP and  supports slow roll models of inflation, the BICEP2 result for $r$ is in some tension with the Planck observations \cite{Planck14} that give $r < 0.11$ and which favour  Starobinsky $R+R^{2}$ inflation~ \cite{Starobinsky80,Starobinsky83} that predicts $r=0.003$.  The BICEP2 value for $r$  is consistent with simple single scalar field $\phi $ models of chaotic inflation with a quadratic self-interaction $V(\phi) =m \phi^{2}/2$~\cite{Linde1985} that predict $r \simeq  0.16$.   Although aspects of the BICEP2 analysis require further investigation, if the BICEP2 result for $r$ is confirmed then it is of interest to note that such a value follows naturally from RSII braneworld cosmology. Slow roll chaotic inflation in RSII braneworld  cosmologies for a single field with a monomial potential $V(\phi )=V_{0}\phi^{p}$  has been the subject of several  investigations. Although early studies~ \cite{Maartens2000,Langois2000} found that the ratio of tensor-to-scalar modes was suppressed for inflation at high energy scales $\rho \gg \sigma$,  subsequent investigations~\cite{Liddle2003,Tsujikawa2004a,Tsujikawa2004b,Zarrouki2011,Calcagni2013} found instead an enhancement. Specifically, the spectral index, $n_s$, and tensor-to-scalar ratio, $r$, in the RSII braneworld scenario are given respectively by
\begin{align}
n_s - 1 &= -\frac{2(2p + 1)}{N(p + 2)},\nonumber\\
r &= \frac{24p}{N(p+2)}, \quad \rho \gg \sigma ,\nonumber
\end{align}
where $N$ is the number of $e$-folds\footnote{Slightly different forms of the denominator in the expression for $r$ are given in the literature. For example, \cite{Liddle2003} give $N(p+2)+p-1$  whereas~\cite{Zarrouki2011,Calcagni2013} give  $N(p+2)+p$.}. For a quadratic potential and $N \approx 55-60$, the values of $n_s$ and $r$ agree closely with the BICEP2 result.

\end{document}